\documentstyle[preprint,aps]{revtex}
\begin{document}
%\tightenlines
\preprint{APCTP-1999006// hep-th/9903058}
\title{\Large \bf AdS$_3$ Black Hole Entropy \\ and the Spectral Flow on the Horizon}
\author{Jin-Ho Cho$^{1,2}$
\thanks{jhcho@galatica.snu.ac.kr}
and Soonkeon Nam$^{1,2}$
\thanks{nam@nms.kyunghee.ac.kr}
}
\address{$^1$
Department of Physics\\
and Basic Sciences Research Institute\\
 Kyung Hee University, Seoul, 130-701, Korea \\
}
\address{$^2$APCTP, Seoul, 130-012,  Korea\\ 
}

\maketitle
\begin{abstract}
We consider the entropy problem of $AdS_3$ black holes using the conformal 
field theory at {\it the horizon}. We observe that the supersymmetry 
is enhanced at the  horizon of massless $AdS_3$ black hole. This allows us 
to determine the vacuum of the modular invariant conformal field theory to 
be the NS-ground state (which corresponds to $AdS_3$ spacetime). This is 
smoothly related to the R-ground state (corresponding to massless black hole) 
by a spectral flow, which can be understood as a superconformal transformation. 
\end{abstract}
\tightenlines
%
%\section{Introduction}
\baselineskip=24pt
The microscopic origin of the entropy of black holes has been 
a challenging problem in quantum gravity since its original
formulation\cite{BH}. It has proven to be a fertile ground to test the 
ideas of string theory\cite{string}, 
and the conformal field theory on D-branes\cite{Vafa,DbraneBH} 
has been quite successful.
However, in the D-brane approach, the geometric picture was not so
clear because it was formulated in the weak coupling limit.
The fact that it can be extended to near-extremal
cases\cite{nearextreme} has been taken with a grain of salt, since the 
D-branes are BPS objects. 
Better understanding of the Bekenstein-Hawking entropy could follow
from the relationship of the BTZ black hole\cite{BTZ}
in 2+1 dimensions and  higher dimensional black 
holes in string theory.
This possibility is due to the observation that the near horizon geometry of 
higher dimensional black hole configurations can be related to that of 
BTZ black hole by some duality transformations\cite{Hyun}.
This makes the study of BTZ black hole quite important.
Since black holes in 2+1 dimensions have asymptotic geometry of 
anti-de Sitter ($AdS$) space, some kind of holographic principle might 
hold a key to the problem of black hole entropy, in light of recent 
developments in $AdS$/CFT\cite{Maldacena}.
However, at least for the case of BTZ black hole, we need not resort 
to the full string theory on $AdS_3\times S^3 \times M^4$ to solve the 
entropy problem.

In this paper, we consider the entropy problem of $AdS_3$ black holes using 
the conformal field theory at the horizon (to be precise we mean the 
apparent horizon). First we observe that the supersymmetry is enhanced 
at the  horizon of massless $AdS_3$ black hole. This allows us to 
determine the vacuum of the modular invariant conformal field theory
to be the NS-ground state (which corresponds to $AdS_3$ spacetime). This 
is smoothly related to the R-ground state (corresponding to massless 
black hole) by a spectral flow, which can be understood as a 
superconformal transformation. 

Let us now describe the BTZ black hole\cite{BTZ}.
There are no curvature singularities for the solution so that the BTZ
black hole solution came as a surprise. 
The metric of black hole of mass $M$ and angular momentum $J$ is
\begin{equation}
ds^2 = - N^2 dt^2 + N^{-2} dr^2 + r^2 (d\phi+N^{\phi} dt)^2 ,
\end{equation}
where the lapse and shift functions are
\begin{equation}
N =  \left( -8GM+\frac{r^2}{l^2} + \frac{16G^2J^2}{r^2}   \right)^{1/2},
\ \ \  N^\phi = -\frac{4GJ}{r^2}.
\end{equation}
The asymptotic symmetry of $AdS_3$ is generated by two copies of
the Virasoro algebra with generators $L_n, \  n\in{\rm integer}$, with 
central charge $c= 3l/2G$\cite{Brown}.
Although the bulk degrees of freedom are nondynamical,
we have a nontrivial dynamical conformal field theory (CFT) on the boundary.
Such conformal symmetry is also found on the horizon of the black 
hole\cite{Carlip}.

Let us sketch the current status of study on BTZ black hole.
Applying the Cardy's formula\cite{Cardy} for the asymptotic growth of 
states for  CFT,  
one can count the number density of the microscopic degrees.
Using this Carlip obtained the black hole entropy from CFT 
on {\it the  horizon}\cite{Carlip}. 
The derivation was based on the fact that gravity in
$2+1$ dimensions can be formulated as topological Chern-Simons 
theory\cite{Achucarro}, and the boundary dynamics at the horizon is a 
CFT described by the $SL(2,R)\times SL(2,R)$ Wess-Zumino-Witten model.
However, the difficulties with this approach were spelled out 
recently\cite{Carlip3}. Carlip's original work \cite{Carlip} is 
flawed by the fact although he resorts to large $k$ limit, 
where $k$ is the level number of the $SL(2,R)$ Kac-Moody algebra, 
$k$ actually is not so large, on shell, at the point $\hbar r_- \sim r_+ k$ 
which is the most relevant to the black hole entropy in his calculation. 
This was remedied with a simpler boundary condition for the  horizon, 
and he obtained the central charge without the details of the boundary 
CFT, and derived entropies even for higher dimensional black 
holes\cite{Carlipnew}. Similar results were also found by 
Solodukhin\cite{solo}.

On the other hand, Strominger obtained the entropy from the CFT at the 
{\it asymptotic boundary} of $AdS_3$ black hole\cite{Strominger}.
This very simple and elegant observation, however,  cannot tell what 
the boundary degree of freedom is, and whether the central charge is 
the {\it effective} central charge $c_{\rm eff} = c- 24\Delta_{\rm min}$ 
or not, just as in the case of Carlip's approach\cite{Carlipnew}. 
(See also \cite{solo}.) Here $\Delta_{\rm min}$ is minimum value of the 
conformal dimension of the CFT. We will elaborate on this point later.

Applying the Regge-Teitelboim method\cite{Regge},
Ba\~nados et al. obtained the algebra satisfied by the global 
charges without the details of boundary theoryi \cite{Banados}. 
There is also an explicit derivation using a boundary system coupled to the 
bulk geometric background\cite{Taejin}.
However, as is noted by Carlip\cite{Carlip3}, the central charge is modified 
to be $c_{\rm eff} =1$ and is too small to account for the black hole entropy.

There have been many related works due to the recent 
keen interest of $AdS$/CFT duality of Maldacena\cite{Maldacena}.
For example the works of Martinec\cite{Martinec} takes the $AdS$/CFT
seriously and argues that Liouville field theory (derived from Chern-Simons 
Gravity) is just an effective theory corresponding to the macroscopic 
description, and cannot account for the black hole
entropy. There is also a stringy interpretation of the entropy\cite{Giveon}.
Despite these conformal field theoretic approaches, it has already been 
pointed out that none is completely satisfactory\cite{Carlip3}.

In this paper we revisit the CFT approach to the black hole entropy problem.
In deriving the Cardy's formula used to calculate 
the entropy, the following two ingredient are quite essential.
First,  the partition function of a CFT must have modular invariance, 
$\tau\rightarrow -1/\tau$, where $\tau$ is the modular parameter.
Secondly, to evaluate the partition function using saddle point 
approximation, the value of the central charge has to be shifted to 
$c_{\rm eff} = c-24\Delta_{\rm min} $, whenever the ground state eigenvalue 
$\Delta_{\rm min}$ of $L_0$ does not vanish\cite{Kutasov}.  Here we stress 
that $\Delta_{\rm min}$ should be evaluated on {\it the plane},
i.e. $L_0$ is the zero mode of the stress energy tensor on the plane, even
though the partition function is that of a CFT on a cylinder\cite{Cardy}.

Getting the correct number for $\Delta_{\rm min}$ is the source of the 
difficulty in the problem. Extracting the exact information on the 
conformal data ($c$, $\Delta_{\rm min}$)  of the CFT 
starting from a black hole geometry is usually quite difficult.
One way to obtain $\Delta_{\rm min}$ easily is to make use of supersymmetry. 
In the Neveu-Schwarz(NS)-sector of a superconformal field theory, 
$\Delta_{\rm min} = 0$ always, and Ramond(R)-sector has value of $c/24$.
$AdS_3$ geometry is identified as the bosonic backgrounds of (1,1)-type $AdS$ 
supergravity\cite{Coussaert}.
To be more specific, we have the following cases:
(i) the $AdS_3$ vacuum(global $AdS_3$ spacetime) has four Killing spinors 
which are antiperiodic (NS-sector),
(ii) a massless black black hole has two periodic Killing spinors (R-sector),
(iii) a massive extremal black hole has one periodic Killing spinor (R-sector).
This can be easily seen by analyzing the Killing spinor equations for 
each geometry:
\begin{equation}
D_\lambda\chi=\frac{\epsilon}{2l}\gamma_\lambda\chi.
\end{equation}
In the above $D_\lambda=\partial_\lambda  + \frac{1}{4}\omega^{ab}_\lambda 
\gamma_a\gamma_b$ is the covariant differential form with respect to the 
spin connection $\omega^{ab}$, $\epsilon=\pm1$ and 
$\{\gamma_a,\gamma_b\}=2\eta_{ab}$. Two values of $\epsilon$ is possible 
because there are two independent representations
for Clifford algebra in three spacetime dimension.
Naively one expects that the massless black hole corresponds to the ground 
state of the boundary CFT, and this looks reasonable because $AdS$ vacuum 
($M=-1/8G$) is disconnected from the black-hole spectrum ($M\geq 0$),
although $AdS$ vacuum has the lowest energy.

However, this choice of the ground state gives a wrong answer for the 
Bekenstein-Hawking entropy, while using the $AdS$ vacuum instead gives 
the correct answer. If we restrict the spectrum solely to the black holes, 
the corresponding CFT seems to be restricted to the R-sector only.
We know very well that we cannot have a modular invariant theory restricting
to the R-sector only, because the contribution to the partition function 
from the R-sector transforms into that of NS-sector
under some modular transformations\cite{Cappelli}.

Furthermore the operator product expansion (OPE) algebra for superconformal 
theory is such that
\begin{equation}
 [R]\times[R]\sim [NS], \ \  [NS]\times[NS]\sim [NS],
\ \ [R]\times[NS]\sim [R].
\end{equation}
This means that we cannot restrict the CFT to the R-sector only. Moreover, 
the identity will come out from the OPE algebra eventually. If we examine 
this more carefully, we note the $AdS$ vacuum is not quite disjoint from 
the black hole spectrum, but is connected via singular point particle 
geometries between them (see preprint version, hep-th/9204099 of \cite{BTZ}).
This becomes quite clear as we consider the boundary geometries.

Actually one can create massless black hole out of $AdS_3$ by head-on 
collision of two massless particles\cite{Matschull}.
In this case mass gap, $\Delta M =1/8G$, 
between the $AdS_3$ spacetime and massless black hole exactly matches with
the energy of those two particles. The spatial geometry of a point 
particle has a conical singularity\cite{Deser}.
Here the conical singularity is not a serious problem. One can
resolve the singularity by replacing the matter distribution over
a small region for the point particle. In fact, in the high energy
scale, this point particle structure can be resolved. The most 
important thing is that the string cannot see this orbifold fixed point
as singular.

Now we look into the boundary conformal structure starting from its geometry.
The metric on the $r=r_0$ surface for the geometry generated by the point 
particle source of mass $m=-\alpha^2/8G$ in the bulk \footnote{here we 
have negative mass because we set the massless black hole case to have 
$m=0$} is
\begin{eqnarray}
ds^2=r_0^2\left(-{dt^2 \over l^2}+d\phi^2\right),
\qquad\phi\sim\phi+2\pi\alpha, \qquad 0 < \alpha \leq 1,
\end{eqnarray}
from which we note the deficit angle $2\pi(1-\alpha)$.
In the above we have redefined $(1+l^2/r^2_0)^{1/2} t$ as $t$.
It is  convenient to  work in the Euclidean scheme.
\begin{equation}
ds_E^2=r^2_0\left(d\tau^2+d\phi^2\right),
\end{equation}
where $\tau=it /l$ is the Euclidean time. One can focus on
the $r=r_0$ region by  rescaling 
$d\tilde{s}^2\equiv ds^2/r^2_0=d\tau^2+d\phi^2$. 
Although the boundary topology is $S^1\times\Re$, the angular geometry has a
deficit angle. 
We can map this cylindrical geometry with deficit angle to 
a conical one, i.e. the boundary conformal cone,
by the following exponential conformal mapping 
$w=e^{\tau+i\phi}(\bar{w}=e^{\tau-i\phi})$:
\begin{equation}
ds_{\rm cone}^2=dwd\bar{w}\equiv R^{2\alpha-2}(dR^2+R^2d\theta^2),
\end{equation}
where  $w$ is the holomorphic coordinate on the cone. 
In the above we have introduced the polar coordinates
$R=(\alpha e^\tau)^{1/\alpha}$ and $\theta=\phi/\alpha$ to show the conical 
structure, where $0\leq R<\infty,\quad\theta\sim\theta+2\pi$.
With another conformal transformation 
$w(z)=z^\alpha/\alpha,\quad\bar{w}=\bar{z}^\alpha/\alpha$
we get to the conformal plane:
\begin{eqnarray}
ds_{cone}^2=z^{\alpha-1}\bar{z}^{\alpha-1}dzd\bar{z}.
\end{eqnarray}

So $\alpha=1$, which is the plane geometry, 
corresponds to the AdS vacuum and the singular limit $\alpha\rightarrow 0$ 
approaches to the massless black hole. The mass parameter of the point 
particle is a continuous parameter which interpolates these two limits.

The stress energy tensor of the CFT on the cone is obtained from the 
following conformal transformation with the Schwarzian
derivative:
\begin{eqnarray}
T_{\rm plane}(z)\rightarrow z^{2(\alpha-1)}T_{\rm cone}(w)-{c \over 24z^2}(\alpha-1)(\alpha+1).
\end{eqnarray}

Thus the conformal weight of a primary on the cone is related to that
on the plane as follows (with similar expressions for $\bar{L}_0$):
\begin{eqnarray}
(L_{\rm cone})_0=(L_{\rm plane})_0+{c \over 24}(\alpha-1)(\alpha+1).
\end{eqnarray}
Here we note that $L_{\rm cone}$ interpolates $L_{\rm cylinder}$ 
(for $\alpha\rightarrow 0$) and $L_{\rm plane}$ (for $\alpha=1$).
For spinning particle case, we can follow similar step. In this case $L_0$ 
and $\bar{L}_0$ is shifted differently to have $L_0 - \bar{L}_0\neq 0$.

There is another thing which connects these two limit. This is the spectral flow between these
two states in the corresponding boundary CFT.
However, to realize spectral flow one actually needs an extended supersymmetry. 
In fact, one can show that at the black hole horizon, there is supersymmetry 
enhancement. To see this let us consider the Killing spinor equation for 
the massless black hole case.
\begin{eqnarray}
D\chi&=&\left(d+{1 \over 2}{r \over l}\left({dt \over l}\gamma_0\gamma_1-
d\phi\gamma_1\gamma_2\right)\right)\chi\nonumber\\
&=&{\epsilon \over 2l}\left({rdt \over l}\gamma_0
+{l dr\over r}\gamma_1+rd\phi\gamma_2\right)\chi
\end{eqnarray}

The solutions are
\begin{eqnarray}
\epsilon=1;\qquad\chi&=&\sqrt{r}\chi_+ \nonumber\\
\chi&=&\left({1 \over \sqrt{r} }
+{\sqrt{r}  \over 2l}x^\alpha\gamma_\alpha\right)\chi_-\\
\epsilon=-1;\qquad\chi&=&\left({1 \over \sqrt{r} }
-{\sqrt{r}  \over 2l}x^\alpha\gamma_\alpha\right)\chi_+\nonumber\\
\chi&=&\sqrt{r}\chi_-, 
\end{eqnarray}
where  $x^\alpha\gamma_\alpha={t}\gamma_0/l+\phi\gamma_2$.
Due to the $\phi$-dependent terms, only two out of these four Killing spinors 
survive upon the identification $\phi\sim\phi+2\pi$. 
However near the horizon the $\phi$-dependent terms drop out because they are
relatively small compared to the other $1/\sqrt{r}$ terms, and all four 
solutions survive enhancing supersymmetry. Such an enhancement is not a 
surprising thing as we can see in similar cases discussed before\cite{enhance}. 
Actually this enhancement of supersymmetry at the boundary CFT is consistent 
with the spacetime supersymmetry. So if the  horizon is where the boundary 
CFT (for the microscopic degrees of freedom for BTZ black holes) is located, 
then we can solve the entropy problem.
The (1,1)-type $AdS$ supergravity in the bulk gives rise to 
(2,2) supersymmetric horizon CFT. 

One can ask, when the mass of the black hole is zero, if the 
boundary CFT makes any sense, because the  horizon is just a point. It is 
not a problem because the BTZ black hole coordinate patch cannot cover 
the whole region of $AdS_3$. In fact, $r=0$ `point' is the null
surfaces in the global coordinates for $AdS_3$, deliminating the 
Poincar\'{e} region\cite{BHTZ}.
Below the black hole spectrum, that is to say, for the $AdS_3$ spacetime 
and conical geometries of point
particles, one can take the boundary at any point with finite radius.
This is so because the boundary geometry looks the same regardless of 
the value of the radius we have.

The isomorphism, which maps the R-sector to the NS-sector in the (2,2) 
supersymmetric CFT is the spectral flow\cite{Schwimmer}.
We want to show that the spectral flow in fact is a symmetry 
transformation, i.e. superconformal transformation in superspace.
To see this we write down the super stress-energy tensor in 
superspace formalism as follows:
\begin{eqnarray}
{\cal J}(z, \theta^+,\theta^-)=J(z)+\theta^+G^-(z)+\theta_-G^+(z)+i\theta^+\theta^-T(z).
\end{eqnarray}
The superconformal transformation of the stress energy tensor is given by
\begin{eqnarray}
{\cal J}(z,\theta^+,\theta^-)\rightarrow
({\cal D}^+\tilde{\theta}^-{\cal D}^-\tilde{\theta}^+)\tilde{\cal J}(\tilde{z},\tilde{\theta}^+,\tilde{\theta}_-)
+{ik \over 4}S(Z,\tilde{Z}),
\end{eqnarray}
where ${\cal D}^\pm={\partial \over\partial \theta^\mp}+\theta^\pm{\partial \over \partial z}$ 
are the superderivatives and
%\begin{equation}
%S(Z,\tilde{Z})={\partial {\cal D}^+\tilde{\theta}^-\over {\cal D}^-\tilde{\theta}^+}
%-{\partial {\cal D}^-\tilde{\theta}^+\over {\cal D}^+\tilde{\theta}^-}
%-2{\partial\tilde{\theta}^+\partial\tilde{\theta}^- \over {\cal D}^+\tilde{\theta}^-{\cal D}^-\tilde{\theta}^+}
%\end{equation}
$S(Z,\tilde{Z})$ is the $N=2$ super-Schwarzian derivative\cite{Cohn}. $Z=(z,\theta^+, \theta^-)$ is the complex
$N=2$ supercoordinate. 
In general, the superconformal transformation on a super Riemann surface is 
given as
\begin{eqnarray}
\tilde{z}=f(z),\quad\tilde{\theta}^+=\mu^{++}(z)\theta^++\mu^{+-}(z)\theta^-,
\quad\tilde{\theta}^-=\mu^{-+}(z)\theta^++\mu^{--}(z)\theta^-,
\end{eqnarray}
with the following superconformal condition:
\begin{eqnarray}
\mu^{++}=\sqrt{f'}e^{-i\xi}, \qquad\mu^{--}=\sqrt{f'}e^{i\xi},
\qquad\mu^{+-}=\mu^{-+}=0,  
\end{eqnarray}
for some functions $f(z)$ and $\xi(z)$. Conventional conformal transformation
corresponds to the case of $\xi=0$.
 The prime denotes the derivative with respect to $z$.
Under this the components of super stress-energy tensor transform as
\begin{eqnarray}
J(z)&\rightarrow&f'\tilde{J}(\tilde{z})-{k \over 2}\xi'\nonumber\\
G^+(z)&\rightarrow&(f')^{3 \over 2}e^{i\xi}\tilde{G}^+(\tilde{z}),
\qquad G^-(z)\rightarrow(f')^{3 \over 2}e^{-i\xi}\tilde{G}^-(\tilde{z})
\nonumber\\
T(z)&\rightarrow&(f')^2\tilde{T}(\tilde{z})+2\xi'f'\tilde{J}(\tilde{z})
+{k \over 4}\left[-2(\xi')^2+{f''' \over f'}
-{3 \over 2}\left({f'' \over f'}\right)^2\right].
\end{eqnarray}
Denoting the right hand sides as $J_\xi(z),G^\pm_\xi(z)$ and $T_\xi(z)$ 
respectively, one can simplify the whole expression as
\begin{eqnarray}
&&J_\xi(z)=J_0(z)-{c \over 6}\xi',\quad G^\pm_\xi(z)
=G^\pm_0(z)e^{\pm i\xi(z)},\\ \nonumber 
&&T_\xi(z)=T_0(z)+2\xi'J_0(z)-{c \over 6}(\xi')^2,
\end{eqnarray}
where $k=c/3$ was used. We note here $J_0(z),G^\pm_0(z)$ and $T_0(z)$ are 
nothing but the standard superconformal transformation. Rephrasing $J$ as 
$iJ/2$ and $\xi$ as $-i\eta\ln{z}$, we are led to the spectral flow map for 
the $N=2$ superconformal symmetry. 
This means that the spectral flow can be understood as a kind of 
superconformal transformation. The `twist' operator which connects the 
NS-vacuum and R-vacuum is not just a tool here but is in the set of 
primary operators. (The same interpretation is expected for the 
spectral flow for the higher extended superconformal symmetry.) The true 
vacuum turns out to be the NS-vacuum. This makes $c_{\rm eff}=c$, whatever
theory the horizon CFT is.

Another advantage of the spectral flow is that we can understand the the 
black hole creation (R-state) from particle (NS-state) collisions in terms 
of OPE. The particle collision would be OPE of two NS-states, and it seems 
that making R-state is impossible. However with the spectral flow, we can 
actually make
\begin{equation}
[NS]_\eta \times [NS]_{\eta'} \sim [NS]_{\eta+\eta'}
\end{equation}
If $\eta + \eta' = 1$ actually $[NS]_1 = [R]$. 
The Cardy's formula, which counts the physical degrees of freedom, must be 
invariant under the symmetry, and thus under the spectral flow.
Relevance of spectral flow for black hole entropy was discussed in 
ref.\cite{Martinec2}, where the unitary representation of $N=2$ superconformal 
algebra\cite{Nam} is used. Description of the spectral flow in
terms of charged particles coupled to the Chern-Simons gauge theory was 
discussed in different setting\cite{Kogan}. 

To calculate the black hole entropy 
from this horizon CFT, we have to know the central charge $c$ and the 
conformal weight $\Delta$ of the horizon state of the black hole.
We use the results recently obtained by Carlip\cite{Carlipnew}.
\begin{equation}\label{data}
c = \frac{3r_+\beta}{GT},\qquad \Delta = \frac{r_+T}{8G\beta},
\end{equation}
where $\beta$ is the inverse Hawking temperature and $T$ is an 
arbitrary periodicity, which does not affect the result for the entropy.
Differently from the asymptotic CFT, the central charge $c$ depends on the 
inner and outer horizon radius $r_-,\,\,r_+$ of the black hole. 
Different black hole gives different CFT on the horizon. Another 
remarkable thing for the horizon CFT is that the central charge $\bar{c}$ 
of the right moving mode vanishes in the case. The same features were
also found in \cite{solo}.

In three dimension, one can determine the arbitrary periodicity $T$ as follows. 
Given a black hole of mass $\hat{M}$ and angular momentum $\hat{J}$, the conformal 
data are fixed as in (\ref{data}). Each horizon state of conformal weight 
$(\delta,\bar{\delta})$ contributes to the bulk mass $M$ 
and angular momentum $J$. 
Although the energy of a horizon state needs not be equal to the bulk mass 
they must be proportional to each other;
$\delta+\bar{\delta}=\gamma Ml+\zeta$, where $\gamma$ and $\zeta$ are 
dimensionless constants to adjust the different scalings and different 
base points of the energy respectively. One can also find the relation 
between the angular momentum of the horizon state and the bulk
angular momentum $J$ as $\delta-\bar{\delta}=\gamma J$.
The energy gap beween the R-vacuum and NS-vacuum of the horizon CFT matches 
with the mass gap between the massless black hole and the $AdS$ vacuum.
\begin{eqnarray}
{c \over 24}={\gamma l \over 8G},
\end{eqnarray}
which tells us $c=3\gamma l/G$ and $\beta/T=\gamma l/r_+$. Therefore the 
ambiguity of the periodicity $T$ is concerned with the different energy
scalings of the horizon CFT and the bulk $AdS$ geometry. From Carlip's
results and some known facts about the $AdS$ vacuum one can determine 
$\gamma$ and $\zeta$ completely, therefore fix the periodicity.
\begin{eqnarray}
&&\delta(M,J)={r_+ \over 2l}\left(Ml+J+{l\over 8G} 
\right){\beta\over T}\nonumber\\
&&\bar{\delta}(M,J)={r_+ \over 2l}\left(Ml-J+{l\over 8G} 
\right){\beta\over T}\nonumber\\
&&{\beta \over T}={\sqrt{2}l 
\over\left(l^2+\left(r_++r_-\right)^2\right)^{1/2}}.
\end{eqnarray}
The central charge $c$ together with the conformal weight 
$\delta(\hat{M},\hat{J})=\Delta$ results in the correct
statistical entropy $S\sim2\pi r_+/4G$.

Lastly, we would like to comment on the success of
the string theory which gives the correct entropy, 
regardless of the delicate arguments we had to give in 
supergravity theory.
The successes which specify the microscopic structure are
those works making use of the BPS arguments. At least in the 
weak coupling regime, they point out that the microscopic structure
resides on the world volume of D-branes. In this weak coupling
region, it is meaningless to say the bulk geometry. For the
D1-D5-KK case, the effective world volume theory is the (4,4) 
supersymmetric sigma model on the symmetric product of K3\cite{Vafa}. 
Owing to the extended supersymmetry, the Cardy's formula can be applied
at any point of ground state under the spectral flow as long as
both R-sector and NS-sector are fully included in the calculation.
Fortunately for this world volume theory, there is no reason to
pick out the R sector ground state as the true ground state.

There are several works which seem to produce the correct central charge 
and entropy using the  AdS/CFT correspondence,
also without referring to the points we have resolved above.
In fact, one needs to address the same question of the true ground state 
for this scheme also so far as the microscopic structure is not specified.
(Even though the microscopic structure is not specified, one can
determine the ground state using supersymmetry.) In the calculation
of the two point function of stress energy tensor, it has been 
a priori assumed that the ground state is the NS-sector since the 
Poincar\'{e} coordinates (without identification) are usually used.
However, if one neglect the 3-sphere part of $AdS_3\times S_3$, the 
supersymmetry on the asymptotic boundary is just $N=1$ for the 
R sector ground state because there is no supersymmetry enhancement on the
asymptotic boundary differently from the horizon. Therefore spectral
flow is not expected in this asymptotic CFT. One has to resort to other 
method to say that the NS vacuum is the true vacuum of the asymptotic CFT. 

{\bf Acknowledgements}
We would like to thank J. de Boer, A. Giveon and N. Ohta for discussions.
This work is supported by KOSEF (981-0201-002-2) and by Korea Research Foundation (1998-015-D00073).

\end{document}